\documentclass{kapproc}

\normallatexbib
\def\be{\begin{equation}}
\def\ee{\end{equation}}
\def\bea{\begin{eqnarray}}
\def\eea{\end{eqnarray}}

\begin{document}
\input epsf

\articletitle{The field-to-particle transition problem}

\author{Jer\'onimo Cortez}
\affil{Instituto de Ciencias Nucleares, Universidad Nacional Aut\'onoma de M\'exico\\
 A. P. 70-543 M\'exico D.F. 04510}
\email{cortez@nuclecu.unam.mx}

\author{Leonardo Pati\~no}
\affil{Instituto de F\'\i sica, Universidad Nacional Aut\'onoma de M\'exico\\
 A. P. 20-364,  M\'exico D. F. 01000}
\email{leonardo@ft.ifisicacu.unam.mx}

\author{Hernando Quevedo}
\affil{Instituto de Ciencias Nucleares, Universidad Nacional Aut\'onoma de M\'exico\\
 A. P. 70-543 M\'exico D.F. 04510}
\email{quevedo@nuclecu.unam.mx}

\chaptitlerunninghead{The field-to-particle transition problem}

\begin{abstract}

We formulate in an intuitive manner several conceptual aspects
of the field-to-particle transition problem which intends to extract
physical properties of elementary particles from specific field configurations.
We discuss the possibility of using the conceptual basis of the holographic
principle and the mathematical fundaments of nonlinear sigma
models for the field-to-particle transition. It is shown that
certain classical
gravitational configurations in vacuum
may contain physical parameters with discrete values, and that
they behave under rotations as particle-like objects.

\end{abstract}


\section*{Introduction}

Probably, the oldest dream of many generations of physicists has been
to find a consistent and simple way for describing the phenomena
we observe in nature. Simplicity is in this context an important feature.
In all our research proposals we intent to extract the simplest aspects of a
phenomenon and to put them as the fundaments for constructing a consistent
theory. It is in this connection that physicists today argue
that all phenomena in nature should be described by only four
types of interactions. One important goal of today's research is to
show that in fact we are dealing with probably only one interaction
which manifests itself differently at different levels. If this
turns out to be true, we will be left with one fundamental interaction
which will be responsible for all the physical processes.

General relativity and quantum mechanics can be considered as the
most important conceptual cornerstones in the development of physics
during the last century. General relativity gave us the possibility to
understand the physical phenomena related with the gravitational
interaction at the classical level, whereas quantum mechanics gave rise
to the development of quantum field theories which describe the fundaments
of the weak and strong interactions. The standard model which includes
also the electromagnetic interaction constitutes the basis of modern
elementary particle physics.

However, in our opinion, in the process of development of the modern
physical theories we have been mixing different concepts. On the one
hand, we know from experience that two ingredients are necessary in
order to construct a realistic model of a physical interaction, namely,
fields and matter. In gravity, for example, when we want to describe
the gravitational field of a matter distribution, we take an action with a
term which contains the field itself and a second term which represents
matter. Whereas we know (more or less)
from physical and mathematical arguments how
to select the correct term that plays the role of field, the term
corresponding to matter has to be imposed {\it ad hoc}. Matter is
an external entity that has to be postulated in accordance with
our understanding of its specific properties. The question arises,
do we really need to postulate matter in such a rough manner?
Of course, we cannot argue that this approach is not correct. In fact,
it is one of the best we know, the standard model of elementary particles
being an example of its great success. But we can argue that this method
is not simple. If our goal is to simplify the way we do physics, we
are going in the wrong direction. It is difficult to imagine that
at the end of the job, we will have one fundamental interaction
with all possible types of matter and all possible types
of different properties. Instead, it would be more convenient to end
with one fundamental interaction which gives rise (through a variational
principle) to many special field configurations, each of them corresponding
to a different particle or constituent of matter. This dream is known
as the field-to-particle transition problem.

It is the aim of this work to present several intuitive ideas about
the methods one could apply in order to attack the field-to-particle
transition problem. We believe that it is necessary to start with the
analysis of old and new ideas in which the tendency to simplicity is
inherent. In Section II we will briefly review the main aspects of the
recently formulated holographic principle. We will see that the intrinsic
aim of the holographic principle is to replace complex physical systems
by other simpler systems without loosing information. Section III is
devoted to a review of nonlinear sigma models from which we expect
to obtain some of the mathematical tools necessary in order to
compare different theories in different spaces. A concrete example
of classical gravitational field configurations from which particle-like properties can
be derived is presented in Section III. We end with
a discussion in Section IV.

\section{The holographic principle}

The conceptual origin of the holographic principle was settled in the 70's by
Bekenstein \cite{bek} and Hawking \cite{haw}, who formulated the second law of
thermodynamics for black holes and the process of particle creation by black holes, respectively.
These ideas were implemented in the context of
quantum field theories, specially in quantum gravity and cosmology by
t'Hooft \cite{thooft} and Susskind \cite{suss}.

The holographic principle is a statement about
the counting of the quantum states of a physical system. Let us begin with some
intuitive ideas that should clarify the meaning of this statement.
Consider a region $B$ of
space. No condition is imposed on the topology or geometry of the region
$B$, but for the sake of simplicity one can begin by identifying $B$
with a sphere. Then, let $V$ be the volume of that sphere. Suppose that
we introduce in the interior of $B$ an arbitrary physical system so that
the region exterior to $B$ is empty. Now consider the space ${\cal H}$ of states
that describe the physical system inside $B$. One can then ask the question
about the dimensionality of the state space ${\cal H}$.
Obviously, the answer will
depend on the physical characteristics of the system. Suppose that it consists of
 a lattice of spins with spacing $d$. If the lattice fills entirely the region $B$,
  then the maximum number of spins contained in it is $V/d^3$.
Since each spin can be in two different states, the total number of
orthogonal states is $2^{V/d^3}$ and this is also the dimensionality of the
state space ${\cal H}$.
This number also determines the maximum entropy $S_{max}$
of the system which is defined as the logarithm of the total number of
states, $S_{max}=\ln N_{states} = \ln {\rm dim} ({\cal H})$.
So, in this example we have that $S_{max}=(V\ln 2)/d^3$.
This counting process and its relation to the maximum entropy of a system
is what holography intents to do in a general setting. In our example,
no further information can be extracted because we are dealing with a
system with a finite number of degrees of freedom and, consequently, the
dimensionality of ${\cal H}$ (Hilbert space) is finite.

The interesting cases are those
in which the number of degrees of freedom is infinite as in field theory.
To handle this case one has to determine the entropy density $s$ as
a function of the energy density $\rho$ of the field. Then $S_{max}=
s(\rho_{max}) V$ and the total number of states is $N_{states}=\exp (S_{max})$.
Now suppose that in the region $B$ we have a set of fields sources including
gravity. Let $A$ be the area of the boundary $\partial B$. The maximum mass
of the system contained within $\partial B$ cannot exceed the mass of black
hole of horizon area $A$. This is a crucial point. According to our theoretical
understanding of field theories, we do not know of any physical system
that, being localized within a certain region, could possess a mass
greater than that of black hole whose horizon area coincides with the area
of the boundary of that region. In other words, black holes are the most
massive objects in nature. Now, we know from the second law of black hole
thermodynamics that the maximum entropy of a black hole of area $A$ is
$S_{max} = A/4$ (we use Planck units with $c=G=\hbar = 1$). Consequently,
the total number of states has the bound $N_{states}=\exp(A/4)$. This is
the statement of the holographic principle in this particular case.
This simple relationship has deep implications. It relates a quantity
in the bulk ($N_{states})$ with a different quantity on the boundary ($A)$.
On the other hand, it predicts an upper bound for the dimensionality of
the state space ${\cal H}$. Notice that we did not impose any conditions on the
fields and matter distribution which fills $B$. That is, the fields can
be classical or quantum. Both cases have been analyzed in the literature.

In the above description we did not specify the region $B$ and we have
freely ``inserted" fields into it. However, it is well known that
especially relativistic fields affect the topology and geometry of
the space where they live. Therefore, one of the main challenges when
trying to develop specific examples is to define the region $B$ in
accordance with the existing fields. In the case of cosmological models,
this has been done for the Friedmann-Robertson-Walker (FRW) spacetime
\cite{fissus} and some classical generalizations \cite{bou}. It has
been found that the dynamics of the FRW spacetime ``in the bulk" is
governed by an entropy relationship ``on the boundary". This result
has been generalized to include different types of quantum corrections
\cite{quantum}.  For the case where the fields in $B$ are quantum,
the most studied example is that of string theory in $B= AdS_5 \otimes S_5$
which turned out to be completely equivalent to a Super-Yang-Mills theory
on the boundary of the AdS space $\partial B = \partial AdS$ \cite{malda}.

We now turn back to the field-to-particle transition problem.
First of all, let us mention that this problem is within the conceptual
idea of the holographic principle. The internal properties of the field (in
the bulk) should become represented by external properties of the
particle (on the boundary). Let $\varphi_0$ be a specific field
configuration, which is solution of a classical field theory, that
describes the spin of a particle. The total number of states is
$N_{states} = 2$. If we now fill the region where the particle lives
with a black hole, then, according with the above discussion, the
``area" of the particle is $A=4\ln 2$. Up to here, nothing especial
seems to happen. However, when we try to analyze the dynamics of this
system, we find the problem of the zero-modes \cite{raja}.
The fluctuations
of the field $\varphi_0 +\delta \varphi_0$
lead to unstable configurations for the elementary particle.
This contradicts our daily experience since elementary particles are stable
with respect to infinitesimal perturbations. This problem is due to
the fact that fluctuations occur in the field, which has an infinite number
of degrees of freedom, and affect the particle, a system with a finite
number of degrees of freedom. To handle this problem properly and
in accordance to the conceptual idea of the holographic principle,
one has to follow the following steps: (i) Select the field $\varphi$,
together with its underlying theory, and a specific field configuration
$\varphi_0$ that describes the spin of the particle;
(ii) Define the region $B$ in accordance with the geometrical properties
of $\varphi_0$; (iii) Find the theory on $\partial B$ and the configuration
$\tilde{\varphi}_0$ which are the counterparts  of the theory in $B$ and
the field $\varphi_0$, respectively;
(iv) Perform a perturbation around the new
specific configuration $\tilde{\varphi}_0+\delta \tilde{\varphi}_0$.
If the entire procedure works correctly, it could be expected that the
configuration ``on the boundary", $\tilde{\varphi}_0$, is stable.
At the moment, all this is just a speculation. However, we will see
in Section III that in the framework of Einstein's gravity  one can
find a large class of gravitational configurations that show
a spin-like behavior.

We think that one of the main obstacles in performing the above
procedure is that no exact mathematical tools are known for
``going from the bulk to the boundary". In the next Section, we
will  review a mathematical construction that could help
to a better understanding of the difficulties.

\section{Nonlinear Sigma Models}

Nonlinear sigma models (NLSM) are an important theoretical laboratory in the framework
of field theory, particularly those defined on Riemannian symmetric spaces, which are
the integrability condition of the classical theory \cite{A-A-R} (they admit an infinite
 number of conservation laws and are examples of completely integrable field theories).
  Nonlinear sigma models are, in several ways, closely related to Yang-Mills theories
   and they have points of resemblance with QCD \cite{qcd-res} (in $1+1$ spacetime,
   some of these models have asymptotic freedom \cite{poly} and instantons \cite{instantons}).
    The Einstein equations for some gravitational fields, for instance axisymmetric fields,
     Einstein-Rosen gravitational waves, $T^{3}$ and $S^{1}\times S^{2}$ Gowdy
     cosmological models, etc., are related to the equations for nonlinear models in $1+1$
 spacetime \cite{varios}. NLSM also appear in string theory and the similarities between the
 gravitational and sigma fields \cite{dewitt} are enough to make them of interest as toy models
  for quantum gravity.

Roughly speaking, a NLSM is a field theory of maps between manifolds with the
following properties: (a) the fields are subject to nonlinear constraints
and (b) the Lagrangian density and the constraints are invariant under the
action of a global symmetry (Lie) group $G$. More precisely \cite{bala,A-A-R},
the classical field configurations in such a model are maps $\phi:B\to M$,
 where $B$ is a given base space and $M$ is a given target space.
 The description ``nonlinear'' is reserved to those models where the
  physical fields for all points $p \in B$ take values in a (Riemannian)
  manifold $M$ which is not a linear space. This restriction is to guarantee
  positivity of the energy in the corresponding NLSM. Since any Riemannian
  manifold $M$ can be isometrically embedded into a Euclidean vector space $E$,
   the Lagrangian density of the model (rewritten in terms of $E$-valued fields)
    has to be supplemented by the constraints expressing the fact that the $E$-valued
    fields must be restricted to lie on the embedded submanifold $M$.

In most of these models the global invariance group $G$ acts transitively on $M$;
 here we shall assume that this is indeed the case (thus, $M$ is a Riemannian
 homogeneous space for $G$). If $H$ is the stability group of a point $m \in M$,
 then $M$ can be identified with the space of left cosets $G/H$ (i.e., $M=\{gH\}, \ g\in G$).

General methods exist for constructing Lagrangians for these theories, the idea
is simply to represent the field configurations of the model not by maps $\phi$
 from $B$ to $M$ but by maps $g$ from $B$ to $G$, with $\phi (x)=g(x)H$. Now,
 the Lagrangian density ${\cal{L}}$ in any nonlinear model is a function of $g$
 and $\partial_{\mu}g$, ${\cal{L}}={\cal{L}}(g, \partial_{\mu}g)$, that is
  invariant under the gauge transformation
\begin{equation}
\label{gauge-tran}
g(x) \mapsto g(x)h(x)\: , \:\:\: h(x) \in H \ ,
\end{equation}
in such a way that the gauge invariant fields have values in $G/H$ and the
Lagrangian density can be regarded as a function of fields with values in $G/H$.

The construction of ${\cal{L}}$ proceeds as follows. Let $G$ be a faithful
representation of a compact semisimple Lie group $G$ (global symmetry group)
 and let $\{b(\rho)\}$ be a basis for the Lie algebra ${\cal{G}}$ of $G$ with
 the following property: ${\rm{Tr}}(b(\rho)b(\sigma))=\delta_{\rho , \sigma}$,
 where $\rho , \sigma =\{1,...,[G]:={\rm{dim}}\,G\}$. For $\alpha \leq [H]:={\rm{dim}}\,H$,
  the generators $b(\alpha)$ are taken to span the Lie algebra ${\cal{H}}$ of $H$ and
   we denote them by $t(\alpha)$. The remaining generators are called $s(i)$
    with $[H]+1 \leq i \leq [G]$. Thus the commutation relations are
\begin{equation}
\label{cr}
[t(\alpha),t(\beta)]=i\,f_{\alpha \beta \gamma} t(\gamma)\:\: ,
\end{equation}
\begin{equation}
[t(\alpha),s(i)]=i\,\bar{f}_{\alpha ij} s(j)
\:\: ,
\end{equation}
\begin{equation}
[s(i),s(j)]=i(\bar{f}_{\alpha ij}t(\alpha)+f_{kij}s(k)) \ .
\end{equation}
Let $\omega$ be the one-form defined on $G$ with components
$\omega_{\mu}(g)=g^{-1}\partial_{\mu}g$ which under a gauge
transformation of the form (\ref{gauge-tran}) transforms as
\begin{equation}
\label{w-tran}
\omega_{\mu}(gh)=h^{-1}\omega_{\mu}(g)h+h^{-1}\partial_{\mu}h \ .
\end{equation}
It is not difficult to see that one can write $\omega_{\mu}$ as
 a sum of two terms (actually, projections of $\omega_{\mu}$ into
  the Lie algebra ${\cal{H}}$ and its orthogonal complement):
\begin{equation}
\omega_{\mu}=A_{\mu}+B_{\mu} \ ,
\end{equation}
where $A_{\mu}(g)=t(\alpha){\rm{Tr}}(t(\alpha)\omega_{\mu}(g))$ and
 $B_{\mu}(g)=s(i){\rm{Tr}}(s(i)\omega_{\mu}(g))$. Accordingly, under
  a gauge transformation these components behave as
\begin{equation}
\label{AB-tran}
A_{\mu}(gh)=h^{-1}A_{\mu}(g)h+h^{-1}\partial_{\mu}h \:\: , \:\:\:\:\:\: B_{\mu}(gh)=h^{-1}B_{\mu}(g)h \ .
\end{equation}

It is worth noticing that we have the structure of a principal fiber
bundle $({\bf{E}},{\bf{B}},{\bf{F}},{\bf{G}},{\bf{\Pi}})$ with total
space ${\bf{E}}=G$, base space ${\bf{B}}=G/H$, fiber ${\bf{F}}\simeq H$,
structure group ${\bf{G}}=H$ and projector ${\bf{\Pi}}:G\to G/H$, $g \mapsto [g]$. By construction all fields lie in $G$ and the physical fields lie in the base space $G/H$. Due to this structure we have then that the one-form $A$ with components $A_{\mu}$ transforms like a gauge potential (for the gauge group $H$) and therefore it acts as a connection one-form. Thus, from $A$ we can construct the curvature two-form $F$ which under (\ref{gauge-tran}) transforms as $F'=h^{-1}Fh$.

The structure of the one-form $\omega$ allows us to construct different
quantities satisfying the above given requirements to be Lagrangian densities
for a NLSM. In particular,
\begin{equation}
\label{f-density}
{\cal{L}}=\sqrt{-g}g^{\mu \nu} {\rm {Tr}} (B_{\mu}B_{\nu}) \ ,
\end{equation}
and
\begin{equation}
\label{s-density}
{\cal{L}}=\sqrt{-g}g^{\mu \tau}g^{\nu \rho} {\rm {Tr}} (F_{\mu \nu}F_{\tau} \rho) \ ,
\end{equation}
where $g^{\mu \nu}$ are the components of the given metric tensor on
 (the spacetime, for example) $B$, in (local) coordinates $x^{\mu}$,
 and $g$ its determinant.

If $\Phi$ is a gauge invariant field, then we can express it as follows
\begin{equation}
\label{adj}
\Phi=g\bigl( {\textstyle{\sum_{\alpha}}}t(\alpha)\bigr)g^{-1} \ .
\end{equation}
Since the r.h.s of (\ref{adj}) belongs to ${\cal{G}}$, then
\begin{equation}
\label{lincomb}
g\bigl( {\textstyle{\sum_{\alpha}}}t(\alpha)\bigr)g^{-1}={\textstyle{\sum_{\rho}}}\phi_{\rho}b(\rho) \ ,
\end{equation}
where the fields $\phi_{\rho}$ are also physical fields. From Eq.(\ref{lincomb})
and the normalization property for the generators of ${\cal{G}}$, one can show
that the following relationship  must hold
\begin{equation}
\label{cons}
{\textstyle{\sum_{\alpha}}}{\rm{Tr}}(t(\alpha)t(\alpha))={\textstyle{\sum_{\rho}}}\phi^{2}_{\rho}{\rm{Tr}}(b(\rho)b(\rho)) \ .
\end{equation}
It follows from Eq.(\ref{cons}) that the fields $\phi_{\rho}$ are subject to
 a nonlinear constraint and there are $[G]-1$ independent physical fields.
 The manifold $M$ is defined by Eq.(\ref{cons}) and, due to the nonlinear
character of this constraint, is not a vector space.

An interesting case of a NLSM is the so-called harmonic map. Considering
$x^\mu$ as the local coordinates in $B$, the fields $\phi_b$ define the original
map $\phi : B \rightarrow M$ which is called harmonic if the corresponding
Lagrangian (\ref{f-density}) or (\ref{s-density}) satisfies a minimum action
principle. This condition leads to a set of (partial) differential equations
in $M$ that, in general, can be made to be related to the field equations
of the fields in $B$. In particular, when the field $g_{\mu\nu}$ in $B$ is
taken to satisfy Einstein's vacuum equations with two Killing vector fields,
the fields $\phi_b$ in $M$ turn out to satisfy a geodesic equation. Thus, a
gravitational field (with the appropriate symmetry) in $B$ is represented
by a geodesic in $M$. This also corresponds to a reduction of the number
of degrees of freedom, at least at the level of differential equations,
and, accordingly, to a ``projection"  of a theory in the bulk $B$ to an
equivalent theory on the boundary $M$.

In the context of holography we can proceed as follows. Suppose we have
a field living in $B$ and we want to ``project" it into its boundary
$\partial B$. If we intend to apply a harmonic map for this projection,
we need to look at the case in which the target space $M$
is the boundary of the base space $B$, {\it i.e.},
$\phi: B \rightarrow \partial B$. This is allowed in the construction
explained above. The next step is the identification of the target space
$\partial B$ with the space of left cosets $G/H$ and the construction
of the corresponding Lagrangian density. The details of this construction
will vary, depending on the original field and on $B$ itself. In general,
one could expect some arbitrariness in the choice of the space of left
cosets since for a given boundary space $\partial B$ the choice of the
group $G$ that acts transitively on it is not unique. Each selection would
lead, in general, to a different Lagrangian density and hence to a different
theory on the boundary. It is not clear under which criteria the selection
of the ``right" theory should be done. One possibility could be to impose
some relationships between the field equations in $B$ and the corresponding
equations in $\partial B$.

It is interesting to note that in general a harmonic
map does not impose any connection between the base and target spaces.
This allows us to analyze specific examples of the holographic principle
in which the boundary is not really a boundary in the topological sense.
For instance, in the case of string theory in $AdS_5\otimes S_5$ one
might think that its boundary is a (8+1)-dimensional space. However,
its boundary turns out to be a (3+1) space that coincides with the
boundary of the AdS spacetime. All these kinds of possibilities are
allowed in the context of harmonic maps.

The intuitive ideas described above about the implementability of the
holographic principle and nonlinear sigma models in the context of
the field-to-particle transition problem are all based upon the assumption
that there are field configurations that emulate physical properties
of elementary particles. In the next Section we will show that in fact
such configurations exist at the level of classical fields.


\section{The field-to-particle transition problem in gravity}

An interesting possibility from the point of view of gravity is to think of an
 elementary particle as a specific gravitational configuration. The conceptual
idea was first proposed by Wheeler~\cite{Wheeler}, who introduced the term
``geon" as an abbreviation for ``gravitational-electromagnetic entity", an
electromagnetic field configuration which would keep together by its own gravitational
 interaction having the approximate properties of a particle.
In a general sense, a geon is considered today as a special gravitational
configuration with a nontrivial topology.

A topological geon as a gravitational
field configuration is easy to construct; difficulties
arise when we try to obtain a topological geon which reproduces
 the behavior of a particle. To this end, we could use the great
 richness of the field of exact solutions to Einstein-Maxwell equations;
 however, the problem is that there is not a recipe to construct such a
 geon and, of course,  it is not appropriate to apply the method of trial
 and error with all the possible spacetimes. The real task is to find generic
 conditions which should be satisfied by spacetimes which are supposed to
 represent a particle, and then to restrict the analysis only to the subset
  of solutions to Einstein-Maxwell equations which satisfy those conditions.

One of the first differences that comes to mind between a particle and a
gravitational configuration is that the intrinsic properties of a particle,
such as mass, charge, and spin are quantized, wile the parameters involved
in a gravitational configuration can take values in a continuum. In view of
this fact, the consequent question is, how can a configuration with continuous
 parameters account for an entity with discrete parameters? This is precisely
  the question we have addressed in a previous work~\cite{pq}, and we will
  briefly review our arguments here.

There are different approaches to answer this question. Here, our proposal is
to extend Dirac's argument about the quantization of the electric charge~\cite{Dirac}
 to the case of gravitational configurations. Remember that the quantum phase
  acquired by a charged particle describing a closed path in the presence of
  a magnetic field can be computed by means of the expression
\begin{equation}
\Phi =e^{{iq}\int _{s}B},           \label{dp}
\end{equation}
where $B$ is the magnetic field, $q$ the charge of the particle
 traveling along a closed loop, and $s$ a surface having the path
 of the particle as its boundary.
It is possible to use different surfaces in Eq.(\ref{dp}) for the same path,
 and as long as the surfaces are homotopic, the resulting phases will be identically the same.
  Dirac used the fact that the form of the field of a magnetic monopole allows
   the construction of non homotopic surfaces with the same boundary, and then,
    by requiring that the phases computed using two such surfaces be equal, he
    got to quantize the electric charge.

The key point to notice is that the integral in Eq.(\ref{dp}) is in fact the
 integral of the electromagnetic tensor $F_{\alpha \beta}$ over a spatial surface.
  Furthermore, from the geometrical point of view of field theory, the electromagnetic
   tensor is just the curvature associated to the electromagnetic connection. This gives
    us a clue about how to extend the argument to gravitational configurations.
    Consider a pseudo-Riemannian manifold $(M,g_{\mu \nu})$, where $g_{\mu \nu}$ is
    the underlying metric. Then we can introduce a phase-like object
\begin{equation}
\Phi=e^{\int R},   \label{3}
\end{equation}
where $R$ is the curvature associated with the metric, namely the Riemann tensor,
 which is an endomorphism valued two-form (see, for instance,~\cite{Baez}). Now
 we will use this object to repeat Dirac's argument
in the case of gravity.

The first problem we have is that the components of the Riemann tensor, when understood as
an endomorphism valued two-form, are endomorphisms, so they live in the tangent and
cotangent spaces to the manifold at the specific point of evaluation. If we perform
the integral in (\ref{3}) just as it appears, we will be adding objects which live
in different spaces and this summation will not be justified. What we need is to
 perform a parallel translation of the endomorphism from the point of evaluation
 to a specific point where we will say that the integral is based.
So instead of  Eq.(\ref{3}) we will use
\begin{equation}
\Phi=e^{\int H^{-1}R H},  \label{4}
\end{equation}
where $H$ is the holonomy resulting from the parallel transport that depends explicitly
on the point of evaluation.

Of course there are infinite different paths along which the parallel transport can be done.
 Therefore, the specific form of  $H$ will have to be fixed by using physical arguments~\cite{pq2}.
  For the moment the important point is that there are some conclusions that can be stated
  independently of the explicit form of  $H$; the only thing we will require is that it
  preserves some fundamental symmetries of the curvature. Since $H$ is directly related
  with the metric and its symmetries, this behavior is not unnatural to be expected.

To repeat Dirac's argument we need to use spacetimes which allow the existence of non
 homotopic surfaces, for instance, spacetimes with curvature singularities. Moreover,
  we will restrict our study to vacuum solutions to Einstein equations. For the sake
  of generality, we consider the Petrov classification~\cite{Petrov} of the curvature
   tensor, and analyze each Petrov type separately.
Since the components of the Riemann tensor are endomorphisms which map a four dimensional
space into itself, they must have four eigenvalues
$\lambda_i \ (i=1,2,3,4)$, and it can be
shown~\cite{pq} that for all Petrov types these eigenvalues satisfy the relationships
\begin{equation}
\lambda _{1}=-\lambda _{2} \qquad {\rm and}\qquad  \lambda _{3}=-\lambda _{4}. \label{lambdas}
\end{equation}
Furthermore, there are ways to guarantee~\cite{pq,pq2} that the property (\ref{lambdas})
is preserved by the eingenvalues of the endomorphism obtained by  the integral in Eq.(\ref{4}).

In terms of the eigenvalues of the integral in (\ref{4}) the phase obtained can be expressed as
\begin{equation}
\Phi = T
\left( \begin{array}{cccc}
e^{\lambda_{1}} & 0 & 0 & 0 \\
0 & e^{-\lambda_{1}} & 0 & 0 \\
0 & 0 & e^{\lambda_{3}} & 0 \\
0 & 0 & 0 & e^{-\lambda_{3}}
\end{array} \right) T^{-1}, \label{fasee}
\end{equation}
where $T$ is a real matrix that describes a coordinate transformation.

Let us now consider spacetimes with metrics that are invariant with respect to the change
$\phi \rightarrow \phi + \pi$ (in spherical coordinates). This condition is satisfied by a
 large class of gravitational configurations, for instance, by all the axially symmetric
 solutions. To perform the calculation of the phase-like object let us consider two non
 homotopic surfaces with common boundary (this is the case, for example, when a curvature
 singularity exists between the two surfaces). Using Dirac's argument about the equality
 of the
phases calculated along the two surfaces, we obtain
the following conditions for the $\lambda$'s
\begin{equation}
\lambda_{1}=i \, n_{1} \, \pi \; \; \; {\rm and}  \;\;\; \lambda_{3}=i \, n_{2} \, \pi , \label{CC}
\end{equation}
where $n_1$ and $n_2$ are arbitrary integers. Since the $\lambda$'s are functions of the
 parameters of the metric, Eqs.(\ref{CC}) can be considered
as relationships similar to those obtained by Dirac. The importance of this
result is that we have reached a discretization in the continuum  of the parameters that
 determine the gravitational configuration. The explicit form of these conditions depend
  on the exact expression for the holonomy $H$. Nevertheless,
the result obtained in Eqs.(\ref{CC}) is quite general because it
does not depend on any explicit value
of the holonomy.

Another interesting result arises when we insert the discrete values (\ref{CC}) in Eq.(\ref{fasee}).
It can be shown that the only possible phases are either  the identity $\Phi = {\bf 1}_{4 \times 4}$,
 or an expression different from the identity but whose square, however, becomes the identity $\Phi ^{2} =
{\bf 1}_{4 \times 4}$.
If we understand the traveling of an observer around an object as an active diffeomorphism, this is
 equivalent to a rotation of the object by $2\pi$. Therefore, saying that the phase acquired by such
  an observer is restricted as we have mentioned above, is equivalent to saying that some of the
  particles we are modeling are invariant under $2\pi$ rotations and others under $4\pi$ rotations.
   This is a surprising result, because the first case corresponds to
the behavior of a boson, and the second one to a  fermion, being the only options allowed. At
 the beginning we were not looking for this prediction, but it arose in a natural way when
 extending Dirac's argument to gravity. It clearly provides another important known characteristic
  of elementary particles. If we were to consider strictly the bosonic or fermionic nature of
   these models, we would need to analyze other important conditions such as the stability
of this particle-like behavior under fluctuations of the underlying field, that is, the problem
 of zero-modes mentioned above. A further crucial condition is that imposed by the
 spin-statistics theorem, that is, the behavior of the model under the interchange of
 identical particles according to this perspective.
This has been already studied in the context of geons by using different approaches
(see, for instance~\cite{Sorkin}).


\section{Conclusions}

In this work we have reviewed some of the main conceptual aspects of the
field-to-particle transition problem. The main goal is to find a different
approach to the study of physical systems in which fields and matter are
involved. In this approach, matter should not be an external entity that
enters the theory in an {\it ad hoc} manner. Instead, matter should be
an additional field component that arises as a specific field configuration.
The first step in this approach is to explore the possibility of
reproducing the physical properties of elementary particles from a
field configuration.

We have described some introductory aspects of the holographic principle,
and we have shown that it can conceptually be used to understand the
intrinsic problems of the field-to-particle transition. In particular,
the problem of zero-modes could be investigated by specifying an
equivalent theory in a different space such that the fluctuations
of the field become described by fluctuations of an equivalent entity
with a finite number of degrees of freedom. This would help to
handle the divergences that appear in the zero-modes.

As the mathematical tool to formulate correctly the field-to-particle
transition and the inherent problem of zero-modes, we propose to
use nonlinear sigma models. We have described how harmonic maps
allow us to project a theory from a space to a different theory
in a different space. Although this procedure is quite arbitrary
in general, we see this as an advantage for the formulation
of apparently different theories which can then be analyzed
under additional restrictions in order to find out their
physical equivalence.

As an explicit example for a field-to-particle transition, we have
analyzed certain gravitational field configurations by using
a phase-like object. It was proven that the physical parameters
entering these configurations become discretizated when we
demand that the phase-like object be equal on two non homotopic
surfaces with a common boundary.  Additionally, we saw that
these configurations behave under rotations either as bosons or as
fermions. No other options are allowed! It is interesting that
classical field configurations show the fermionic behavior, a
property which is usually associated with quantum systems.

The proposals presented in this work are all very rough and
have no deep physical explanations. A more detailed investigation
will be necessary in order to formulate them in a more consistent
manner from the physical and mathematical points of view.
 They should be interpreted more as a first attempt to
formulate questions which bother the authors. However, we
consider that these questions have to situated in the conceptual
kernel of most modern field theories.

\begin{acknowledgments}
This work has been supported by DGAPA-UNAM, grant No. 112401,
and CONACYT-Mexico, grant No. 36581.
J. C. was
supported by a CONACYT-UNAM (DGEP) Graduate Fellowship.
L. P. was
supported by a UNAM-DGEP Graduate Fellowship.

\end{acknowledgments}


\begin{chapthebibliography}{1}
\bibitem{bek} J. D. Bekenstein, {\em Lett. Nuovo Cim.} {\bf 4} (1972) 737.
\bibitem{haw} S. Hawking, {\em Commun. Math. Phys.} {\bf 43} (1975) 199.
\bibitem{thooft} G. t' Hooft, {\em Dimensional reduction in quantum gravity},
gr-qc/9310026.
\bibitem{suss} L. Susskind, {\em J.Math.Phys.} {\bf{36}} (1995) 6377.
\bibitem{fissus} W. Fischler and L. Susskind, {\em Holography and cosmology},
hep-th/9806039.
\bibitem{bou} R. Bousso, JHEP 9906 (1999) 028.
\bibitem{quantum} O. Nojiri, S. Odintsov, O. Obregon, H. Quevedo and M. Ryan, {\em Mod.Phys.Lett.} {\bf A16} (2001) 1181.
\bibitem{malda} J. Maldacena, {\em Int.J.Theor.Phys.} {\bf 38} (1999) 1113.
\bibitem{raja} P. Rajaraman, {\em Solitons and instantons} (Noth-Holland Press,
Amsterdam, 1988).

\bibitem{bala}A.P. Balachandran, A. Stern and G. Trahern, {\em Phys.Rev.} {\bf{D19}} (1978) 2416;
 A.P.Balachandran, G.Marmo, B.S.Skagerstam and A.Stern, {\em{Classical Topology and Quantum States}}
 (World Scientific, 1991).
\bibitem{A-A-R}E.Abdalla, M.C.B.Abdalla and K.D.Rothe,
{\em{Non-perturbative methods in 2 dimensional quantum field theory}} (World Scientific, 1991).
\bibitem{qcd-res}W.Marciano and H.Pagels, {\em Phys.Rep.} {\bf{36C}} (1978) 137; F.J.Yndur\'{a}in,
 {\em{Quantum Chromodynamics}} (Springer-Verlag, 1983).
\bibitem{poly}A.M.Polyakov, {\em Phys.Lett.} {\bf{59B}} (1975) 79.
\bibitem{instantons}A.A.Belavin and A.M.Polyakov, {\em JETP Lett.} {\bf{22}} (1975) 245; A.D'Adda,
M.L\"{u}sher and P.Di Vecchia, {\em Nucl.Phys.} {\bf{B146}} (1978) 63.
\bibitem{varios} C.Misner, {\em Phys.Rev.} {\bf{D18}} (1978) 4510; M.Hirayama, H.Chia Tze,
 J.Ishida and T.Kawabe, {\em Phys.Lett.} {\bf{66A}} (1978) 352; A.Ashtekar and V.Husain,
 {\em Int.J.Mod.Phys} {\bf{D7}} (1998) 549; J.Cortez, D.N\'u\~ nez and H.Quevedo,
 {\em Int.J.Theo.Phys.} {\bf{40}} (2001) 251.
\bibitem{dewitt}B.DeWitt, in: {\em{Geometrical and Algebraic Aspects of Nonlinear Field Theory}},
ed.  S. De Filippo, M.Marinaro, G.Marmo and G.Vilasi,  (Elsevier Science Publishers B.V. North-Holland, 1989).


\bibitem{Wheeler}J. A. Wheeler, {\em{Phys. Rev.}} {\bf{97}} (1955) 511;
{\em{Geometrodynamics}} (Academic, New York and London, 1962).

\bibitem{pq}L. Pati\~{n}o and H. Quevedo, {\em{submitted}}.

\bibitem{Dirac} P.A.M. Dirac, {\em{Proc. Roy. Soc.}} {\bf{A133}} (London, 1931) 60.
\bibitem{Baez}J. Baez and J.Muniain, {\em{Gauge Fields, Knots and Gravity}}, (World Scientific, 1994).
\bibitem{pq2} L. Pati\~{n}o and H. Quevedo, {\em{in preparation}}.
\bibitem{Petrov}D. Kramer, H.Stephani, M.MacCallum and E.Herlt,
{\em{Exact Solutions of Einstein's Field Equations}}, (Cambridge University, 1980).
\bibitem{Sorkin}J. Friedman and R. Sorkin, {\em{Phys. Rev. Lett.}} {\bf{44}} (1980) 1100;
 {\em{Gen. Relativ. Gravit.}} {\bf{14}} (1982) 615.




\end{chapthebibliography}
\end{document}